\documentclass[twocolumn,showpacs,preprintnumbers,amsmath,amssymb]{revtex4}
\usepackage{graphicx}
\begin{document}

%
%
\title{Nonadiabatic effect on the quantum heat flux control}  
\author{Chikako Uchiyama}
\affiliation{%
Faculty of Engineering, University of Yamanashi,
4-3-11, Takeda, Kofu, Yamanashi 400-8511, Japan}%
\date{\today} 
\def\ch{{\cal H}}
\def\cg{{\cal G}}
\def\ct{{\cal T}}
\def\cv{{\cal V}}
\def\els{{\cal L}_{S}}
\def\cha{{\cal H}_{S}}
\def\el{{\cal L}_{1,\nu}^{\chi_\nu}}
\def\elc{{\cal L}^{\chi}}
\def\elk{{\cal L}_{0}}
\def\hbar{\mathchar'26\mkern -9muh}
\def\ih{\frac{i}{\hbar}}
\def\v1{\mathbf{1}}
\def\dw{{\dot W}}
\def\iclt{{\cal L}'_{\chi}(t)}
\def\rhov{{\vec \rho}}
\def\xih{{\hat{\xi}}}
\begin{abstract}
We provide a general formula of quantum transfer that includes the non-adiabatic effect under periodic environmental modulation by using full counting statistics in Hilbert-Schmidt space. Applying the formula to an anharmonic junction model that interacts with two bosonic environments within the Markovian approximation, we find that the quantum transfer is divided into the adiabatic (dynamical and geometrical phases) and non-adiabatic contributions. This extension shows the dependence of quantum transfer on the initial condition of the anharmonic junction just before the modulation, as well as the characteristic environmental parameters such as interaction strength and cut-off frequency of spectral density. We show that the non-adiabatic contribution represents the reminiscent effect of past modulation including the transition from the initial condition of the anharmonic junction to a steady state determined by the very beginning of the modulation. This enables us to tune the frequency range of modulation, whereby we can obtain the quantum flux corresponding to the geometrical phase by setting the initial condition of the anharmonic junction. 
\end{abstract}

\pacs{05.60.-k, 05.70.Ln, 44.05.+e}%
\maketitle
\section {Introduction}
Control of quantum heat flux between multiple environments has attracted much attention from scientists as well as engineers \cite{Phon}. One of the most challenging issues is quantum pumping, which aims to find a flux (a directed transfer of quantum particles) via a joint system between two environments under time-dependent modulations that average out the bias during a period. Ren, Li, and H\"anggi \cite{Ren}  describe the transferred heat under the out-of-phase and sufficiently slow (adiabatic) temperature modulations with the geometrical phase, which is based on the Berry phase \cite{Berry}. Similar treatments are used for chemical reaction systems \cite{SN071}, electron pumping \cite{EPT} (which is experimentally demonstrated \cite{EPE}) and discussions on entropy \cite{SH}.

The condition of sufficiently slow (adiabatic) environmental modulations to describe quantum pumping with the geometrical phase allows the relevant system to approach the steady state sufficiently quickly. The steady state is obtained in the Markovian approximation for setting of environmental parameters each time.  In other words, we require the reasonable frequency range of the external modulation to be much smaller than the reciprocal of the relaxation time of the relevant system \(\tau_{R}\). A frequency range of environmental modulations for anharmonic junction systems with \(\tau_{R}\) of a few {\textrm fs} or {\textrm ps} is estimated to be \(\Omega \ll \) 1 {\textrm THz} in Ref. \cite{Ren}.  Now the following questions arise: What happens when the adiabatic condition \(\Omega \ll 1/\tau_{R}\) is not satisfied, and is it possible to find an optimal condition for quantum pumping by adjusting these parameters? We try to answer these questions in this work.

The non-adiabatic effect on quantum pumping has been mainly discussed using the cyclic modulation of an energy level of a quantum dot \cite{Strass,Rey} or molecular system \cite{Segal06}. To the author's knowledge, the effect under external driving of environmental parameters has only been discussed in studies on stochastic entropy production \cite{Esposito07,EB2010}. They find that the stochastic total entropy production is divided into the adiabatic and non-adiabatic contributions, and that setting the initial condition in the non-equilibrium state and external driving cause the non-adiabatic effect on total entropy production. External driving makes the probability distribution deviate from the steady state. Furthermore, they find that the non-adiabatic effect becomes more significant as driving of environments becomes more sudden\cite{Esposito07}. However, because the modulation of environmental parameter is not cyclic in these studies, they do not yield the non-adiabatic effect on quantum pumping.  

In this work, we provide a general expression for the quantum transfer to include the non-adiabatic effect of modulation of environmental parameters using full counting statistics\cite{EHM}. We apply the formula to an anharmonic junction where a two-level system simultaneously interacts with two environments consisting of an infinite number of bosons. To discuss the non-adiabatic effect, we take a piecewise change of environmental temperatures as the modulation protocol. The protocol is very different from the continuous modulation used in conventional studies, but it enables a clear discussion on the relation between \(\tau_{R}\) and \(\tau_{P}(\propto 1/\Omega)\). We can find examples of sudden switching on and off of system-environment interaction in treating a small system such as single ion \cite{Bustamante,Abah}. With these settings, we look for an optimal condition for \(\tau_{P}(\propto 1/\Omega)\) to generate a quantum flux between environments through a relevant two-level system. 

\section{Formulation} 
First, we provide a general expression for the transfer of a quantum particle between environments and a relevant system using full counting statistics \cite{EHM}. Let us consider a system interacting with two environmental systems labeled by \(L\) and \(R\). The total Hamiltonian is \(\ch=\ch_{0}+\ch_{int}\) with
\begin{equation}
\ch_{0}=\ch_{S}+\sum_{\nu=L,R}\ch_{E,\nu},\;\;\; \ch_{int}=\sum_{\nu=L,R}\ch_{1,\nu} ,
\label{eqn:1}
\end{equation}
where \(\ch_{S}\) is the Hamiltonian of the relevant system, \(\ch_{E,\nu}\) is the Hamiltonian of the \(\nu\)th environment with (\(\nu=\)L, or R), and \(\ch_{1,\nu}\) is the interaction Hamiltonian between the \(\nu\)th environment and the relevant system.  

The full counting statistics gives the time evolution of the transfer from the relevant system into an environment (or vice versa) by using the difference between the outcomes of two-point projective measurements of an environmental variable \(Q\). Let us briefly summarize the formalism for full counting statistics developed by Esposito, Harbola, and Mukamel [13]. Denoting the difference for outcomes \(q_{0}\) and \(q_{\tau}\) at \(t=0\) and \(\tau\) as \(\Delta q=q_{\tau}-q_{0}\), information on the transfer between the system and environment is encapsulated in the probability density of the difference \(P_{\tau}(\Delta q)\), which is defined as
\begin{equation}
P_{\tau}(\Delta q)=\sum_{q_{\tau},q_{0}}\delta(q_{\tau}-q_{0}) P[q_{\tau},q_{0}],  \nonumber 
\end{equation}
where \(P[q_{\tau},q_{0}]\) is the joint probability to obtain outcomes \(q_{0}\) at \(t=0\) and \(q_{\tau}\) at \(t=\tau\).  It is defined as
\begin{equation}
P[q_{\tau},q_{0}]={\rm Tr} [{\hat P}_{q_{\tau}} U(\tau,0) {\hat P}_{q_{0}} W(0) {\hat P}_{q_{0}} U^{\dagger}(\tau,0) {\hat P}_{q_{\tau}}], \nonumber 
\end{equation}
where \({\rm Tr}\) is the trace operation over the total system including the relevant system and environment, \({\hat P}_{q_{\tau}}=|q_{\tau} \rangle \langle q_{\tau}|\) means the projective measurement on \(Q(\tau)\), \(U(\tau,0)\) is the time evolution operator for the total system, and \(W(0)\) is the initial condition of the total system. To find the transfered quantity, the cumulant generating function is introduced, \begin{eqnarray}
S_{\tau}(\chi) = \ln \int P_{\tau}(\Delta q) e^{i \chi \Delta q} d\Delta q \nonumber
\end{eqnarray}
where \(\chi\) is called the counting field. We choose \(\chi_{L}\) or \(\chi_{R}\) as \(\chi\), respectively. The \(n\)th cumulant of \(P_{\tau}(\Delta q)\), \(\langle \Delta q^n \rangle_{c}\), is given by the \(n\)th derivative of \(S_{\tau}(\chi)\), \(\langle \Delta q^n \rangle_{c} = \partial^{n} S_{\tau}(\chi)/ \partial (i \chi)^{n}|_{\chi=0}\), which describes the transfer between the system and environment.  Using the definition of \(P[q_{\tau},q_{0}]\), we find \(\int P_{\tau}(\Delta q) e^{i \chi \Delta q} d\Delta q=\sum_{q_{\tau},q_{0}}e^{i \chi (q_{\tau}-q_{0})} P[q_{\tau},q_{0}]\).  A relation \(\sum_{q_{0}}e^{-i \chi q_{0}}{\hat P}_{q_{0}} W(0) {\hat P}_{q_{0}}=e^{-i (\chi/2) Q(0)} W_{0} e^{-i (\chi/2) Q(0)}\) with \(W_{0} \equiv \sum_{q_{0}}{\hat P}_{q_{0}} W(0) {\hat P}_{q_{0}}\) enables us to rewrite the expression of \(S_{\tau}(\chi)\) as 
\begin{equation}
S_{\tau}(\chi)=\ln {\rm Tr}_{{\rm S}} \rho^{\chi}(\tau),
\label{eqn:2}
\end{equation} 
where \({\rm Tr}_{{\rm S}}\) denotes the trace operation over the relevant system, and \(\rho^{\chi}(\tau)\) is the reduced density operator defined as
\begin{equation}
\rho^{\chi}(\tau)={\rm Tr}_{{\rm E}} U_{\chi/2}(\tau,0) W_{0} U_{-\chi/2}(\tau,0) . \nonumber
\end{equation} 
For \({\rm Tr}_{{\rm E}}\), the trace operation is over the environmental variables and \(U_{\chi}(\tau,0)\) is a time evolution operator modified to include the counting field as \(U_{\chi}(\tau,0)=e^{i \chi Q(\tau)} U(\tau,0) e^{-i \chi Q(0)}\).  For the factorized initial condition between the reduced system \(\rho^{\chi}(0)\) and Gibbs states of the environmental systems \(\rho_{{\rm E}}(=\prod_{\nu=L,R} \rho_{{\nu}})\), the time evolution of \(\rho^{\chi}(\tau)\) is obtained as \(\frac{d}{dt} \rho^{\chi}(t) = \xi^{\chi}(t) \rho^{\chi}(t)\)\cite{EHM}, which corresponds to a TCL (time-convolutionless) type of master equation\cite{TCL1,TCL2,US,Breuer,ucom}. 

While the full counting statistics provides us information on the transfer between system and environment, because of the Liouville operators in \(\xi^{\chi}(t)\), it is difficult to find the mathematical structure between the elements of the reduced density operator \(\rho^{\chi}(t)\). Such difficulties are overcome by transforming it into a vector in Hilbert-Schmidt  (H-S) space to find the formal solution as
\begin{equation}
| \rho (\chi,t) \rangle = T_{+} \exp[\int_{0}^{t} dt' \Xi^{\chi}(t')] | \rho (\chi,0) \rangle,
\label{eqn:3}
\end{equation}
where \(| \rho (\chi,t) \rangle \) is a vector consisting of elements of \(\rho^{\chi}(t)\), \(T_{+}\) is the time ordering operation from right to left, and \(\Xi^{\chi}(t) \) is a super matrix form of \(\xi^{\chi}(t)\) in H-S space. The whole information on the reduced dynamics is expressed with the matrix structure of \(\Xi^{\chi}(t)\).  Equation~(\ref{eqn:3}) describes the exact non-Markovian dynamics for a single environmental parameter setting and a factorized initial condition when we include all orders of system-environment interaction.

The time evolution of the first moment is written as \( \langle \Delta q \rangle_{t} = \langle 1 | \frac{\partial}{\partial (i \chi)} \rho (\chi,t) \rangle |_{\chi=0} \), with the trace operation in H-S space defined as \( \langle 1 |\). Because the density operator is a trace-class operator satisfying the relation \({\rm Tr}_{{\rm S}} \rho^{\chi=0}(t) =1\), the state \(\langle 1 | \) is a left eigenstate of \(\Xi^{\chi=0}(t)\) with zero eigenvalue. Using this relation, we find 
\begin{eqnarray}
\langle \Delta q \rangle_{t}=\langle 1 | \int_{0}^{t} dt' [\frac{\partial \Xi^{\chi}(t')}{\partial (i \chi)}]_{\chi=0} \rho (0,t') \rangle ,
\label{eqn:4}
\end{eqnarray}
which describes the time evolution of the first moment for a single setting of the environmental parameters. (Detailed derivation of Eq.~(\ref{eqn:4}) is shown in Appendix A). When we put the counting field between the system and the \(L(R)\)th environment, we choose the variable of the partial derivative \(\chi= \chi_{L(R)}\).

To study quantum pumping, we need to accumulate the transfer of environmental variables under a cyclic change of environmental parameters during a period \(\ct\). In this work, we focus on the step-like change of environmental parameters as in \cite {EB2010,Abah,ratchet,bm}. This corresponds to situations with using thermal light \cite{EB2010} or engineered laser reservoirs \cite{Abah} as environments. Here we assume that we can switch on and off the system-environment interaction instantaneously.

Dividing the period \(\ct\) into \(n\) intervals with \(\delta t (\equiv \ct/n) \) during which the environmental parameters are constant, and defining the \(j\)th interval as \(t_{j-1}  \leq t < t_{j} \) with \(t_{j}=j \; \delta t\) for \(1 \leq j \leq n\), we obtain the accumulated quantity \(J^{\nu}\) between the relevant system and the \(\nu\)-th environment as 
\begin{eqnarray}
J^{\nu}=\sum_{j=1}^{n} \langle \Delta q^{\nu} \rangle^{j}_{\delta t} ,  \label{eqn:5} 
\end{eqnarray}
where \(\langle \Delta q^{\nu} \rangle^{j}_{\delta t}\) denotes the quantity transferred to the \(\nu\)th environment during the \(j\)th interval, which is given by
\begin{eqnarray}
J^{\nu} =  \sum_{j=1}^{n}\langle 1 | \int_{0}^{\delta t} dt' [\frac{\partial \Xi_{j}^{\chi}(t')}{\partial (i \chi_{\nu})}]_{\chi_{\nu}=0} \rho_{j} (0,t') \rangle.
\label{eqn:6}
\end{eqnarray}
where  \(\Xi_{j}^{\chi}(t')\) is the super matrix for the environmental parameter setting over the \(j\)th interval, and we define \(|\rho_{j} (0,t') \rangle = \kappa_{j}^{\chi}(t')\prod_{m=1}^{j-1}\kappa_{m}^{\chi}(\delta t) | \rho_{1}(0,0) \rangle\) with \(\kappa_{j}^{\chi}(t) \equiv T_{+}\exp[\int_{0}^{t} dt' \Xi_{j}^{\chi}(t')]\). Equation~(\ref{eqn:6}) is obtained assuming the system and environment can be factorized at the beginning of each interval and it can describe the non-Markovian dynamics.  A detailed derivation of Eq.~(\ref{eqn:6}) is presented in Appendix A.

We take positive \(J^{\nu}\) as corresponding to the direction of  quantum transfer from the relevant system into the \(\nu\)th environment. With this definition, the transfer from the environment \(L\) into \(R\) via the relevant system occurs if the net transferred quantity satisfies the relation \(J^{R}-J^{L}>0\). We also consider a finite value of the quantity \(J^{R}-{J^{L}}\) to mean successful quantum pumping. Next, we use the obtained formula to discuss the non-adiabatic effect on quantum pumping of bosons for an anharmonic junction system.  

\section{Application} 
We consider a two-level system (or equivalently a \(\frac{1}{2}\) spin) as an anharmonic junction system \cite{Ren,Segal}, which is supposed to interact with two environmental systems \(L\) and \(R\) consisting of an infinite number of bosons. The Hamiltonian is
\begin{eqnarray}
\ch_{S}&=&\sum_{m=0,1} \varepsilon_{m} |m \rangle \langle m| ,\;\; \ch_{E,\nu}=  \sum_{k} \hbar \omega_{k,\nu} b^{\dagger}_{k,\nu}b_{k,\nu}, \nonumber \\ 
\ch_{1,\nu}&=&  X_{\nu} (|0\rangle \langle 1|+|1\rangle \langle 0|),  \;\;\;\;\; (\nu=L, R)
\label{eqn:7}
\end{eqnarray}
where \(|0\rangle\) (\(|1\rangle\)) is the lower (higher) level of the two-level system. In Eq.~(\ref{eqn:7}), we define \(X_{\nu}= \sum_{k} \hbar g_{k,\nu} (b^{\dagger}_{k,\nu}+b_{k,\nu})\), where \(b^{\dagger}_{k,\nu}\) and \(b_{k,\nu}\) are creation and annihilation boson operators of the \(k\)th mode of the \(\nu\)th environment. For this setup, we study boson transfer under cyclic and piecewise modulation of environmental temperatures \(T_{L}\) and \(T_{R}\).

Applying Eq.~(\ref{eqn:7}) to the generalized master equation including the counting field obtained in \cite{EHM}, and transforming it into H-S space for \(|\rho (\chi,t) \rangle =(\rho^{\chi}_{00}(t),\rho^{\chi}_{01}(t), \rho^{\chi}_{10}(t),\rho^{\chi}_{11}(t))^{T}\), we find a concrete expression of \(\Xi^{\chi}(t)\) for the anharmonic junction model, which is derived in Appendix B.  With the supermatrix \(\Xi^{\chi}(t)\), we find that the time evolution of the diagonal elements of the reduced density operator is decoupled from that of the off-diagonal elements, in analogy with the case without the counting field\cite{Breuer}. This simplifies the evaluation of \(\langle \Delta q^{\nu} \rangle^{j}_{\delta t}\) because we need only the elements in \(\Xi_{j}^{\chi}(t)\) and  \(|\rho_{j}(0,t) \rangle\) that correspond to the diagonal elements as \(\rho^{\chi}_{00}(t)\) and \(\rho^{\chi}_{11}(t)\) for the trace operation in Eq.~(\ref{eqn:6}).

To discuss the non-adiabatic effect on quantum pumping and compare with the adiabatic case in Ref. \cite{Ren}, we focus on the case of weak system-environment coupling and the Markovian (long-time) limit by taking the limit \(t \rightarrow \infty\) to the supermatrix \(\Xi_{j}^{\chi}(t)\), as demonstrated in Appendix C.  In this limit, the elements of \(\Xi_{j}^{\chi}(t)\) are time independent during each interval and determined by setting the environmental parameters. According to each environmental setting, only the reduced density operator \(|\rho_{j}(0,t) \rangle\) evolves in time, which gives
\begin{eqnarray}
\langle \Delta q^{\nu} \rangle^{j}_{\delta t} =\hbar \omega_{0} \{ A^{\nu}(j) \int_{0}^{\delta t} dt' \rho_{00}(t') - B^{\nu}(j) \delta t \}, 
\label{eqn:8}
\end{eqnarray}
where  we define \(\omega_{0}=(\varepsilon_{1}-\varepsilon_{0})/\hbar\).  In Eq.(8), \(A^{\nu}(j)\) and \(B^{\nu}(j)\) are defined as \(A^{\nu}(j) =-(k^{\nu}_{d}(j)+k^{\nu}_{u}(j))\) and \(B^{\nu}(j)= -k^{\nu}_{u}(j)\) with \(k^{\nu}_{d}(j)=\Gamma_{\nu} N_{\nu}(j)\) and \(k^{\nu}_{u}(j)=\Gamma_{\nu} (1+ N_{\nu}(j))\). Here we define  \(N_{\nu}(j)=1/(\exp[\hbar \beta_{j}^{\nu} \omega_{0}]-1)\) for the inverse temperature \(\beta_{j}^{\nu}\) of the \(\nu\)th environment during the \(j\)th interval.  We denote \(\Gamma_{\nu} =2 \pi h_{\nu}(\omega_{0})\) as the coupling spectral density, which is determined by the interaction strength between the system and the \(\nu\)th environment using \(h_{\nu}(\omega)\equiv \sum_{k} g_{k,\nu}^2 \delta(\omega-\omega_{k,\nu})\).  \(k^{\nu}_{u}(j)\) and \(k^{\nu}_{d}(j)\) are rate constants that describe the time evolution of the diagonal elements during \(t_{j-1} \leq t < t_{j} \). Here we set \(t_{0}=0\) and \(t_{n}=\ct\). For \(\rho_{00}(t)=\langle 0 | \rho^{\chi=0}(t) |0 \rangle\), we find 
\begin{eqnarray}
{\dot \rho}_{00}(t)=-K_{d}(j) \rho_{00}(t) +K_{u}(j) \rho_{11}(t) \nonumber
\end{eqnarray}
with \(K_{d}(j)=\sum_{\nu} k^{\nu}_{d}(j)\) and \(K_{u}(j)=\sum_{\nu} k^{\nu}_{u}(j)\) .
We find that the solution to the differential equation for \(\rho_{00}(t)\) is
\begin{eqnarray}
\rho_{00}(t)= \rho_{s}(j) + e^{\Lambda (j) t} (\rho_{00}(t_{j-1})-\rho_{s}(j) ), 
\label{eqn:9}
\end{eqnarray}
where we denote \(\rho_{s}(j)=-K_{u}(j) /\Lambda (j)\) with \(\Lambda (j)=-(K_{d}(j)+K_{u}(j))\).  
Using Eqs.~(\ref{eqn:5}), ~(\ref{eqn:7}), and~(\ref{eqn:8}), we find that \(J^{\nu}\) is divided into the adiabatic and non-adiabatic contributions in the form
\begin{eqnarray}
{\tilde J}^{\nu} = \frac{1}{\ct} J^{\nu} =\frac{1}{\ct}( \cg_{ad}^{\nu}+\cg_{nad}^{\nu})= \frac{\hbar \omega_{0}}{\ct} (\cg_{1}^{\nu}+\cg_{2}^{\nu}+\cg_{3}^{\nu}),\;\;
\label{eqn:10}
\end{eqnarray}
where
\begin{eqnarray}
\cg_{1}^{\nu}&=& \sum_{j=1}^{n} (A^{\nu}(j) \rho_{s}(j)-B^{\nu}(j)) \delta t, \label{eqn:11} \\
\cg_{2}^{\nu}&=& \sum_{j=1}^{n-1} \frac{A^{\nu}(j+1)}{\Lambda(j+1)}(\rho_{s}(j+1)-\rho_{s}(j) ), \;\label{eqn:12} \\
\cg_{3}^{\nu}&=& \sum_{j=1}^{n} \phi^{\nu}_{0}(j)+\sum _{j=2}^{n-1} (\rho_{s}(j-1)-\rho_{s}(j)) \psi^{\nu}(j) \nonumber \\
&&\hspace{1cm}+(\rho_{s}(n-1)-\rho_{s}(n))\frac{A^{\nu}(n)}{\Lambda(n)}e^{\Lambda(n) \delta t}, \;\;\;\label{eqn:13}
\end{eqnarray}
with
\begin{eqnarray}
\phi^{\nu}_{0}(j) &=& (\rho_{00}(0)-\rho_{s}(1)) f^{\nu}(1,j),  \label{eqn:14} \\
\psi^{\nu}(j) &=&\frac{A^{\nu}(j)}{\Lambda(j)}e^{\Lambda (j) \delta t} +\sum_{m=j}^{n-1} f^{\nu}(j,m+1), \label{eqn:15} \\
 f^{\nu}(p,q)&=&\frac{A^{\nu}(q)}{\Lambda(q)} e^{\sum _{\kappa=p}^{q-1} \Lambda_{1 }(\kappa) \delta t }  (e^{\Lambda (q) \delta t}-1). \label{eqn:16} 
\end{eqnarray}
The reason for the above division is as follows: When we take the Riemann sum on \(\cg_{1}^{R}\) and \(\cg_{2}^{R}\) by setting \(n \rightarrow \infty\) and \(\delta t \rightarrow 0\), we find that these reduce to the dynamical and geometrical phase obtained in Ref. \cite{Ren}, respectively. This means that the sum of \(\cg_{1}^{\nu}\) and \(\cg_{2}^{\nu}\) corresponds to the adiabatic contribution. This correspondence is obtained by extending the supermatrix \(\Xi^{\chi}_{d,M}\) to describe the continuous driving as treated in Ref. \cite{Ren} and focusing only on the eigenvector of the extended supermatrix corresponding to the steady state as  given in Appendix D. This is consistent with the expression of \(\cg_{3}^{\nu}\), which shows that, when \(\rho_{00}(0)=\rho_{s}(1)\) and the absolute value of \(\Lambda(j) \delta t\) is sufficiently large, we can neglect \(\cg_{3}^{\nu}\). The former condition corresponds to the adiabatic approximation in Ref. \cite{Ren}, where the population of the relevant system instantaneously approaches the steady state for the temperature setting at an initial time. The expression \(\cg_{3}^{\nu}\) shows that the non-adiabatic contribution to the transferred quantity explicitly depends on the initial condition of the relevant system, \(\rho_{00}(0)\).  Moreover, expanding Eq.(13) about \(\delta t\) up to the first order, we find  
that the non-adiabatic effect described in \(\cg_{3}^{\nu}\) shows a correction to both \(\cg_{1}^{\nu}\) and \(\cg_{2}^{\nu}\).

\section{Numerical evaluation} 
We numerically evaluate the averaged transferred quantity \({\hat J}=({\tilde J}^{R}-{\tilde J}^{L})/\hbar \omega_{0}\) using  Eqs.~(\ref{eqn:10})--(\ref{eqn:16}). Defining frequency of the temperature modulation \(T_{\nu}(t)\) as \(\Omega =\frac{2 \pi}{\ct}\) with \(\ct=n \,\delta t\), we obtain the dependence of \({\hat J}\) by decreasing \(\delta t\) while maintaining \(n\) constant; see Fig. 1. We assume that the system-environment coupling is described by the Ohmic spectral density, \(h_{\nu}(\omega)=s_{\nu} \omega \exp[-\omega/\omega_{c,\nu}]\) for \(\nu=L,R\) with coupling strength \(s_{\nu}\) and cut-off frequency \(\omega_{c,\nu}\). We consider a weak and symmetric system-environment coupling by setting \(s_{\nu}=0.01(=s)\) for \(\nu=L,R\). 
\begin{figure}[h]
\begin{center}
\hspace{0cm}
\includegraphics[scale=0.6]{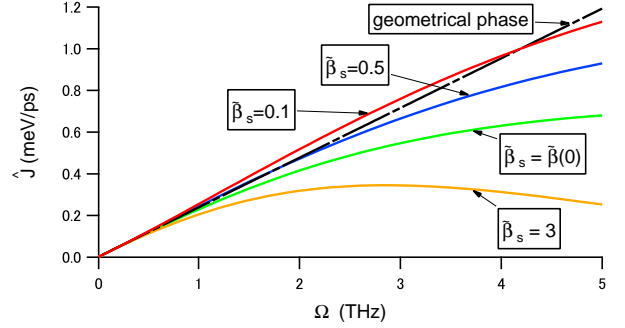}
\end{center}
\caption{(Color online) Frequency dependence for the net current \({\hat J}\) with \(s=0.01\), \(\omega_{c} =3 \omega_{0}\), and \(\hbar \omega_{0}=25 {\rm meV}\) for \({\tilde \beta_{s}}\) values from \(0.1\) to \(3\): (1) the red (light gray) line corresponds to \({\tilde \beta_{s}}=0.1\), (2) the blue (dark gray) line to \({\tilde \beta_{s}}=0.5\), (3) the green (lighter gray) line to \({\tilde \beta_{s}}={\tilde \beta(0)} (\approx 1.07)\), and (4) the orange (lowermost) line to \({\tilde \beta_{s}}=3\). The black dashed line represents the frequency dependence of the net geometrical phase [2]. The temperature modulations, \(T_{L}(t)= 200 + 100 \cos(\omega t+\pi/4)\), and \(T_{R}(t)= 200 + 100 \sin(\omega t+\pi/4)\), are discretized with \(n=41\).}
\label{fig:fig1}
\end{figure}

Since \(\omega_{c,\nu}\) determines the width and peak of \(h_{\nu}(\omega)\) for a constant value of \(s_{\nu}\), the correlation time of the environment, \(\tau_{R}\), becomes shorter as \(\omega_{c,\nu}\) becomes larger. This can be found by studying the time dependence of \(\Lambda \) before taking the Markovian (long-time) approximation \cite{Breuer}, which asymptotically approaches a stationary value more quickly for larger \(\omega_{c,\nu}\). We set \(\omega_{c,\nu} =3 \omega_{0} (=\omega_{c})\), which enables us to consider that \(\Lambda \) coincides with the asymptotic value without the Markovian (long-time) approximation within 5 \(\%\) error at least at the end of each \(\delta t\) for \(5 {\rm THz}\). (This evaluation is done by comparing \(\Lambda \) with the corresponding \(\Lambda(t) \) in the non-Markovian dynamics. A detailed explanation is presented in Appendix E.) This means that the frequency range in Fig. 1 corresponds to the change of time scale \(\tau_{P} (= \delta t)\) from \(\tau_{P} \gg \tau_{R}\) to \(\tau_{P} \approx  \tau_{R}\).  

In Fig.1, we show the frequency dependence of \({\hat J}\) for various initial populations of the relevant system, \(\rho_{00}(0)=\frac{1}{Z} \exp^{-\beta_{s} \ch_{s}}\) with \(Z={\rm Tr}_{{\rm S}} \rho^{\chi=0}(0)\), by setting the scaled quantity \({\tilde \beta_{s}}=\hbar \beta_{s} \omega_{0}\) from \(0.1\) to \(3\), including the case \({\tilde \beta(0)}= \hbar \omega_{0}/ k_{B} T_{\nu}(0) (\approx 1.07)\). We compare these curves with the geometrical phase obtained using Eq.~(\ref{eqn:12}), which is represented as the dashed line in Fig. 1.  For the division number \(n=41\) chosen in this figure, we find that the summation in Eq.~(\ref{eqn:10})  is converged. 
 
 We find that the geometrical phase describes the feature of \({\hat J}\) in the lower frequencies well, but the deviation increases at higher frequencies. We also find that the frequency range where \({\hat J}\) reproduces the value for the geometrical phase becomes larger for smaller \({\tilde \beta}_{s}\), which corresponds to the higher ``temperature" setting of the initial state of the relevant system.  To discuss the dependence of \({\hat J}\) on \({\tilde \beta_{s}}\), we need only the sum of \(\phi^{\nu}_{0}(j)\) over \(j\) as shown in the first term of Eq.~(\ref{eqn:13}).  For our purpose, we set the same situation as \({\hat J}\) on \(\phi^{\nu}_{0}(j)\) by using a quantity that represents the difference between the transfer into the \(R\)th and \(L\)th environment, \({\hat \phi}_{0}(j)=\phi^{R}_{0}(j)-\phi^{L}_{0}(j)\).   In Fig.2, we show the dependence of \({\hat \phi}_{0}(j)\) on \(j\) with changing \(\Omega\) for \({\tilde \beta_{s}}=0.5\). We find that the width of the dependence of \({\hat \phi}_{0}(j)\) on \(j\) becomes larger as \(\Omega\) increases.  This is because an increase in \(\Omega\) means a decrease in \(\delta t\), which causes the value of \(e^{\Lambda (j) \delta t}\) to affect the transferred quantity even for later modulation interval.  This reminiscent effect becomes larger as the ``temperature" setting of the initial state increases, which we can find from the factor \((\rho_{00}(0) - \rho_{s}(1))\) in Eq.~(\ref{eqn:13}).   
\begin{figure}[h]
\begin{center}
\hspace{0cm}
\includegraphics[scale=0.5]{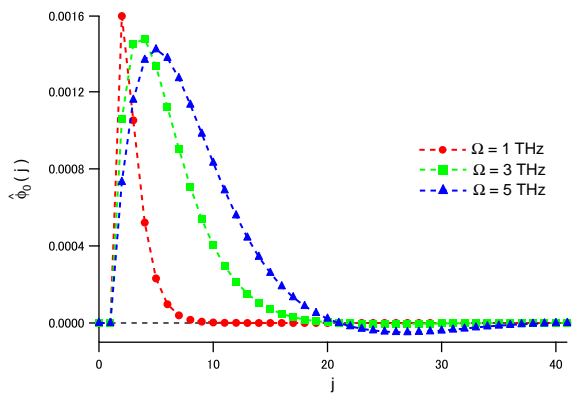}
\end{center}
\caption{(Color online) Dependence of \({\hat \phi}_{0}(j)\) on \(j\) for \({\tilde \beta_{s}}=0.5\) with frequency of the temperature modulation ranging from \(\Omega=1\) to \(5 \,{\rm THz}\): (1) the red dashed line with circles corresponds to \(\Omega=1 \,{\rm THz}\), (2) the green dashed line with squares to \(\Omega=3 \,{\rm THz}\), and (3) the blue dashed line with triangles to \(\Omega=5 \,{\rm THz}\). Other parameters are the same as in Fig. 1. Dashed line is drawn as a guide for the eye.  }
\label{fig:fig2}
\end{figure}

For \({\tilde \beta_{s}} > {\tilde \beta}(0)\), we find \(\rho_{00}(0) < \rho_{s}(1)\) where \({\hat \phi}_{0}(j)\) becomes negative, which decreases the frequency range where we can obtain the contribution of the geometrical phase as shown in the case of \({\tilde \beta_{s}}=3\) in Fig.1.  

Next, let us consider the case, \({\tilde \beta_{s}} = {\tilde \beta}(0)\), where the initial population is already in the steady state under the initial environmental setting.  From Eq.~(\ref{eqn:14}), we find that the initial condition of the relevant system does not affect the pumping current.  From Fig. 1 we find that the non-adiabatic effect on quantum pumping other than the initial condition shows a decrease in the pumping current from the geometrical phase as the modulation frequency increases.

To discuss the feature of quantum pumping, it is necessary to focus on the difference quantity \({\hat J}\) as we have done because the transfer occurs from or into the both environments for general setting of \({\tilde \beta_{s}}\).

\section{Concluding remarks} 
We have provided a general expression for the first moment of quantum transfer under a piecewise modulation of environmental temperatures using full counting statistics. We have applied it to an anharmonic junction model that interacts with two kinds of bosonic environments. The obtained expression includes the non-adiabatic effect under the Markovian approximation, which is an extension of the one expressed with the geometrical phase \cite{Ren}. With this extension, we find a non-linear dependence of net current on frequency of temperature modulation. We also find that quantum pumping depends on the initial condition of the anharmonic junction just before the modulation, as well as the characteristic environmental parameters such as interaction strength and cut-off frequencies of spectral density. For higher initial temperatures of the relevant system, the numerical evaluations show us that the non-adiabatic effect can increase the pumping current over the one obtained under the adiabatic approximation. This means that we can find an optimal condition for the current by adjusting these parameters. We are therefore able to tune the frequency range of modulation to obtain the quantum flux corresponding to the geometrical phase by setting the initial condition of the anharmonic junction.

Although in this study we have focused on the Markovian approximation under weak system-environment coupling, the formula obtained, Eq.~(\ref{eqn:6}), can be used in discussions of the non-Markovian effect on quantum pumping. 
In addition, this formula is applicable to various kinds of physical situations. When applied to the non-equilibrium spin-boson model under a canonical transformation \cite{Ren13,Nicolin}, we can treat the strong system-environment case enabling the heat flux control to be discussed systematically beyond the weak system-environment case treated in this paper. Replacing bosonic environments with fermionic ones and the two-level system with a multi-level system, we can describe the non-adiabatic as well as the non-Markovian effect on electron and/or heat pumping in various systems: (1) We can treat quantum dot systems, which enables us to extend the discussions on adiabatic geometrical pumping in Ref. \cite{EPT}, and (2) applying the formula to metal-molecule-metal systems, we can discuss electron and/or heat pumping in molecular junction, which is attracting intensive theoretical \cite{Phon,Dubi,WYS} and experimental \cite{Chang,Reddy} attention as building blocks of microscopic (thermo)electric devices. We can find concrete experimental examples in a suspended nanotube system between metals \cite{Chang}, and a molecule trapped between a metal substrate and a tip in a scanning tunneling microscope (STM) \cite{Reddy}. Applications to these systems enable us to extend the discussions on thermal rectifications given in \cite{WYS} to provide basic information on the possibility of one or both electron and heat pumping in molecular devices. These remain issues to be treated in the future.

\begin{acknowledgements}
The author thanks Jie Ren, Hisao Hayakawa and Kota Watanabe for their fruitful and helpful comments on this study.  This work is supported by a Grant-in-Aid for Scientific Research (B) (KAKENHI 25287098).
\end{acknowledgements}

\appendix
\section{DERIVATION OF THE FIRST MOMENT OF TRANSFERRED QUANTITY}

In this Appendix, we derive an expression of the first moment of transferred quantity between an environment and a relevant system based on the full counting statistics.  We also derive an expression to describe the accumulated quantity during a period of environmental modulation.

Defining the time ordered exponential in Eq.~(3) as \(\kappa^{\chi}(t) \equiv T_{+}\exp[\int_{0}^{t} dt' \Xi^{\chi}(t')]\), we find the first moment in the form as
\begin{eqnarray}
\langle \Delta q \rangle_{t} &=& \Bigl[\frac{\partial S_{t}(\chi)}{\partial (i \chi)}\Bigl]_{\chi=0} \nonumber \\
&=&   \Bigl[\langle 1 |\frac{\partial \kappa^{\chi}(t)}{\partial (i \chi)}  | \rho (\chi,0)\rangle \Bigl]_{\chi=0} +\Bigl[\langle 1 | \kappa^{\chi}(t) \frac{\partial \rho (\chi,0)}{\partial (i \chi)} \rangle \Bigl]_{\chi=0}. \nonumber \\
\label{eqn:S1}
\end{eqnarray}
Since the reduced density operator \(\rho^{\chi=0}(t)\) is a trace-class operator, a relation \(  \langle 1 | \Xi^{\chi=0}(t) =0 \) is satisfied, which gives 
\begin{equation}
\langle 1 | \kappa^{\chi=0}(t) = \langle 1 |.  
\label{eqn:S2}
\end{equation}
Using these relations, we find 
\begin{eqnarray}
\langle \Delta q \rangle_{t}&=& \Bigl[ \langle 1 |\int_{0}^{t} dt' \frac{\partial \Xi^{\chi}(t')}{\partial (i \chi)} \kappa^{\chi}(t') | \rho (\chi,0)\rangle \Bigl]_{\chi=0},  \;\;
\label{eqn:S3}
\end{eqnarray}
where we used the fact that the second term in the right hand side of Eq.~(\ref{eqn:S1}) does not contribute to \(\langle \Delta q \rangle_{t}\), since the reduced density operator at an initial time,  \( \rho^{\chi}(0)\), does not depend on \(\chi\) when we consider the factorized initial condition for the relevant system and the environmental systems\cite{EHM}.  From these, we obtain Eq.(4).

Next, we consider to accumulate the above quantity during a period of environmental modulation, \(\ct\).  Considering a step-like change of environmental parameter with an interval \(\delta t (\equiv \ct/n)\),  \(| \rho (\chi,\ct) \rangle\) is written as
\begin{equation}
 | \rho (\chi,\ct) \rangle=\prod_{j=1}^{n} \kappa_{j}^{\chi}(\delta t) | \rho (\chi,0) \rangle, 
\label{eqn:S4}
\end{equation}
which gives the accumulated quantity \(J\) in the form as
\begin{eqnarray}
J &=& \Bigl[ \langle 1 | \frac{\partial}{\partial (i \chi)} \rho (\chi,\ct) \rangle \Bigl]_{\chi=0} \nonumber \\
&=&\Bigl[\sum_{j=1}^{n} \langle 1 | \frac{\partial \kappa_{j}^{\chi}(\delta t) }{\partial (i \chi)} \prod_{m=1}^{j-1} \kappa_{m}^{\chi}(\delta t) | \rho (\chi,0) \rangle \Bigl]_{\chi=0}  \nonumber \\
&=& \Bigl[\sum_{j=1}^{n} \langle 1 | \int_{0}^{\delta t} dt' [\frac{\partial \Xi_{j}^{\chi}(t')}{\partial (i \chi)}] \kappa_{j}^{\chi}(t') \prod_{m=1}^{j-1} \kappa_{m}^{\chi}(\delta t) | \rho (\chi,0) \rangle \Bigl]_{\chi=0}. 
\nonumber \\
\label{eqn:S5}
\end{eqnarray}
Defining \(|\rho_{j} (0,t') \rangle = \kappa_{j}^{\chi}(t')\prod_{m=1}^{j-1}\kappa_{m}^{\chi}(\delta t) | \rho (0,0) \rangle\), and setting the variable of partial derivative \(\chi= \chi_{\nu}\) for \(\nu=L\) or \(R\) according to the position of the counting field, we obtain Eq.~(6).

\section{THE SUPERMATRIX FOR THE ANHARMONIC JUNCTION MODEL}
In order to obtain the first moment for the anharmonic junction model, we need to derive the generalized master equation \(\frac{d}{dt} \rho^{\chi}(t) = \xih^{\chi}(t) \rho^{\chi}(t) \), which gives \(\Xi^{\chi}(t)\) and \(\rho(t)\) in Eq.~(6).  In this Appendix, we abbreviate the number of interval \(j\). 

The generalized master equation is derived in \cite{EHM} to give 
\begin{eqnarray}
\xih^{\chi}(t)A &\equiv& -\ih [\ch_{S},A] \nonumber \\
&&\hspace{-1cm} +\sum_{\nu=L,R}  \int_{0}^{t} d\tau {\rm Tr_{E}}[ (i \el(0))(i \el(-\tau)) (\rho_{{\rm E}}A) ], 
\nonumber \\ 
\label{eqn:S6}
\end{eqnarray}
where we assume the factorized initial condition for the relevant system and the environmental systems.  In Eq.~(\ref{eqn:S6}), \(\rho_{{\rm E}}\) denotes the Gibbs state for both environments,  we define \(\chi_{\nu}\) as the counting field between the system and \(\nu\)th environmental system.  \(\el(t)\) in Eq.~(\ref{eqn:S6}) is a modified Liouville operator to include a counting field between the relevant system and the \(\nu\)th environment which is denoted as
\begin{equation}
\el(t) A = \frac{1}{\hbar} [\ch_{1,\nu}^{\chi}(t),A]_{\chi} \equiv\frac{1}{\hbar} [\ch_{1,\nu}^{\chi}(t)A - A\ch_{1,\nu}^{-\chi}(t)], 
\label{eqn:S7}
\end{equation}
for an arbitrary operator \(A\) where  \(\ch_{1,\nu}^{\chi}(t)\) is the Heisenberg picture of \(\ch_{1,\nu}^{\chi}\) which is defined as
\begin{equation}
 \ch_{1,\nu}^{\chi}(t)=e^{(i/\hbar) \ch_{0} t} \ch_{1,\nu}^{\chi} e^{-(i/\hbar) \ch_{0} t} 
\label{eqn:S8}
\end{equation}
with
\begin{equation}
\ch_{1,\nu}^{\chi}=e^{(i/2) \sum_{\nu} \chi_{\nu} \ch_{E,\nu}} \ch_{1,\nu} e^{-(i/2) \sum_{\nu} \chi_{\nu} \ch_{E,\nu}}.  
\label{eqn:S9}
\end{equation}
Replacing \(\el(t)\) with the ordinary Liouville operator, we find that Eq.~(\ref{eqn:S6}) corresponds to the TCL master equation\cite{TCL2,US} by taking up to the second order of the ``ordered" cumulant.  The first ``ordered" cumulant vanishes, since we obtain \({\rm Tr_{E}}  \rho_{E} \ch_{1,\nu}^{\chi}=0\) for the anharmonic junction model whose Hamiltonian is written as Eq.~(7). 

Transforming the reduced density operator \(\rho^{\chi}(t)\) into the vector in the Hilbert-Schmidt space as \(|\rho (\chi,t) \rangle=(\rho^{\chi}_{00}(t),\rho^{\chi}_{01}(t),\rho^{\chi}_{10}(t),\rho^{\chi}_{11}(t))^{T}\), we find that the operator \(\xih^{\chi}(t)\) corresponds to the supermatrix \(\Xi^{\chi}(t)\) in the form as
\begin{eqnarray}
\Xi^{\chi}(t) &=& \Xi_{s}\nonumber \\
&&\hspace{-0.5cm} -\int_{0}^{t} d\tau\left[ {\begin{array}{*{20}{c}}
{{V_{+}}(\tau )}&0&0&{{W_{+}}^{\chi}(\tau )}\\
0&{{Y_{+}}(\tau )}&{Z_{+}^{\chi}(\tau )}&0\\
0&{Z_{-}^{\chi}(\tau )}&{{Y _{-}}(\tau )}&0\\
{{W_{-}}^{\chi}(\tau )}&0&0&{{V_{-}}(\tau )}
\end{array}} \right], 
\nonumber \\
\label{eqn:S10}
\end{eqnarray}
where we define \(\Xi_{s}\) as a diagonal matrix whose diagonal elements are \([0,i \omega_{0}, -i \omega_{0},0]\) and we also define
\begin{eqnarray}
V_{\pm}(\tau )&=& \sum_{\nu=L,R} \{\Phi_{\nu} (\tau ) e^{\mp i \omega_{0} \tau}+\Phi_{\nu}(-\tau ) e^{\pm i \omega_{0} \tau}\}, 
\label{eqn:S11} \\ 
W _{\pm}^{\chi}(\tau )&=& - \sum_{\nu=L,R} \{\Phi_{\nu}(- \hbar \chi_{\nu}- \tau ) e^{\mp i \omega_{0} \tau}  \nonumber \\
&&\hspace{1cm}+\Phi_{\nu}(-\hbar \chi_{\nu} + \tau ) e^{\pm i \omega_{0} \tau}\}, 
\label{eqn:S12} \\
Y_{\pm}(\tau )&=& \sum_{\nu=L,R} 2 \Re (\Phi_{\nu}(\tau )) e^{\mp i \omega_{0} \tau}, 
\label{eqn:S13}  \\
Z_{\pm}^{\chi}(\tau )&=& - \sum_{\nu=L,R}(\Phi_{\nu}(-\hbar \chi_{\nu} - \tau ) +\Phi_{\nu} (-\hbar \chi_{\nu} + \tau )) e^{\pm i \omega_{0} \tau},  
\nonumber \\\label{eqn:S14}  
\end{eqnarray}
with \(\omega_{0}=(\varepsilon_{1}-\varepsilon_{0})/\hbar\), and 
\begin{eqnarray}
\Phi_{\nu}(\tau ) = \sum_{k} g_{k,\nu}^{2}(\langle b^{\dagger}_{k,\nu} b_{k,\nu}\rangle e^{i \omega_{k} \tau} +\langle b_{k,\nu}  b^{\dagger}_{k,\nu}\rangle e^{-i \omega_{k} \tau}).\;\;\;\;\;   
 \label{eqn:S15} 
\end{eqnarray}
Using a continuous spectral density for coupling strength \(g_{k,\nu}\) in Eq.~(\ref{eqn:S15}) as \(h_{\nu}(\omega) \equiv \sum_{k} g_{k,\nu}^2 \delta(\omega-\omega_{k,\nu})\), we obtain
\begin{eqnarray}
\Phi_{\nu}(\tau )&=&\int_{0}^{\infty} d\omega h_{\nu}(\omega) \{n_{\nu}(\omega) e^{i \omega \tau}+(1+n_{\nu}(\omega)) e^{-i \omega \tau} \}  \nonumber \\
&&\hspace{-1cm}=\int_{0}^{\infty} d\omega h_{\nu}(\omega) \{(1+2n_{\nu}(\omega)) \cos(\omega \tau)-i \sin(\omega \tau)\}, 
\nonumber \\ \label{eqn:S16} 
\end{eqnarray}
with \(n_{\nu}(\omega)=1/(e^{\beta_{\nu} \hbar \omega}-1)\).
From Eq.~(\ref{eqn:S10}), we find that the time dependence of diagonal and off-diagonal elements of the density operator are decoupled.  Since Eq.~(\ref{eqn:S3}) means that we need only the diagonal elements of reduced density matrix to obtain \(\langle \Delta q \rangle_{t}\), we pick up the necessary elements from the supermatrix and define it as follows:
\begin{eqnarray}
\Xi^{\chi}_{d}(t) =  - \int_{0}^{t} d\tau \left[ {\begin{array}{*{20}{c}}
{{V_{+}}(\tau )}&{{W_{+}}^{\chi}(\tau )}\\
{{W_{-}}^{\chi}(\tau )}&{{V_{-}}(\tau )}
\end{array}} \right]. 
\label{eqn:S17}
\end{eqnarray}
Substituting \(\Xi^{\chi}_{d}(t)\) into Eq.~(\ref{eqn:S3}), we obtain the first moment of the transferred boson between the two-level system and the \(\nu\)-th environment in the form as
\begin{equation}
\langle \Delta q^{\nu} \rangle_{t} = -\int_{0}^{t} dt' \{w_{+}^{\chi_\nu}(t') \rho_{11}(t') + w_{-}^{\chi_\nu}(t')\rho_{00}(t') \}, 
\label{eqn:S18}
\end{equation}
where we denote \(w_{\pm}^{\chi_\nu}(t)=[\frac{\partial}{\partial( i \chi_\nu)} \int_{0}^{t} d\tau W_{\pm}^{\chi}(\tau)]_{\chi_\nu=0}\). 
Let us note that Eq.~(\ref{eqn:S18}) means the transferred quantity up to \(t\) during which the environmental parameters are set to be constant. 

\section{MARKOVIAN (LONG TIME) LIMIT}
Putting \(t \rightarrow \infty \) on Eq.~(\ref{eqn:S10}), we find the Markovian (long-time) limit of \(\Xi^{\chi}_{d}(t)\) in the form as
\begin{eqnarray}
\Xi^{\chi}_{d,M} \nonumber \\
&&\hspace{-1.5cm}= - \sum_{\nu=L,R} \left[ {\begin{array}{*{20}{c}}
{\Gamma_{\nu} n_{\nu}(\omega_{0})}&{-\Gamma_{\nu} (1+n_{\nu}(\omega_{0})) e^{i \chi_{\nu} \hbar \omega_{0}}}\\
{-\Gamma_{\nu} n_{\nu}(\omega_{0}) e^{-i \chi_{\nu} \hbar \omega_{0}}}&{\Gamma_{\nu} (1+n_{\nu}(\omega_{0}))}
\end{array}} \right], 
\nonumber \\\label{eqn:S19}
\end{eqnarray}
with \(\Gamma_{\nu} \equiv2 \pi h_{\nu}(\omega_{0})\).   Eq.~(\ref{eqn:S19}) describes the time evolution of the two-level system for a single setting of the environmental temperature \(T_{\nu}\) for \(\nu=L,R\).  We find that the elements correspond to the coefficients of the master equation in \cite{Ren}. 
In obtaining  Eq.~(\ref{eqn:S19}), we use relations as
\begin{eqnarray}
\int_{0}^{\infty } \Phi_{\nu} (\tau) e^{-i \omega_{0} \tau} d\tau \nonumber \\
&&\hspace{-3cm}=\int_{0}^{\infty } d\tau \int_{-\infty}^{\infty} d\omega \{h_{\nu}(\omega)n_{\nu}(\omega) \theta(\omega)  \nonumber \\
&&\hspace{-1cm}+h_{\nu}(-\omega)(1+n_{\nu}(-\omega)) \theta(-\omega)\}e^{i \omega \tau}e^{-i \omega_{0} \tau},
\nonumber \\
\label{eqn:S20}
\end{eqnarray}
and
\begin{equation}
\int_{0}^{\infty} e^{i (\omega-\omega_{0}) \tau} d\tau=\pi \delta(\omega-\omega_{0})+i \wp \frac{1}{\omega-\omega_{0}}.
\label{eqn:S21}
\end{equation}
  
Comparing Eq.~(\ref{eqn:S17}) with ~(\ref{eqn:S19}), we find
\begin{eqnarray}
w_{+}^{\chi_\nu}(\infty)&=&[\frac{\partial}{\partial( i \chi_\nu)} \int_{0}^{\infty} d\tau W_{+}^{\chi}(\tau)]_{\chi_\nu=0} \nonumber \\
&&\hspace{-0.5cm} =-\hbar \omega_{0} \Gamma_{\nu} (1+n_{\nu}(\omega_{0})) ,  
\label{eqn:S22} \\
w_{-}^{\chi_\nu}(\infty)&=&[\frac{\partial}{\partial( i \chi_\nu)} \int_{0}^{\infty} d\tau W_{-}^{\chi}(\tau)]_{\chi_\nu=0} \nonumber \\
&&\hspace{-0.5cm}=\hbar \omega_{0} \Gamma_{\nu} n_{\nu}(\omega_{0}) ,  
\label{eqn:S23}
\end{eqnarray}
which gives \(\langle \Delta q^{\nu} \rangle_{t}\) in the Markovian (long-time) limit as
\begin{eqnarray}
\langle \Delta q^{\nu} \rangle_{t} &=& -\int_{0}^{t} dt' \{w_{+}^{\chi_\nu}(\infty) \rho_{11}(t') + w_{-}^{\chi_\nu}(\infty) \rho_{00}(t') \} \nonumber \\
&&\hspace{-0.5cm} = \hbar \omega_{0} \{\int_{0}^{t} dt' \{(-\Gamma_{\nu} (1+2 n_{\nu}(\omega_{0}))) \rho_{00}(t') \nonumber \\
&&\hspace{0.5cm}-(-\Gamma_{\nu} (1+n_{\nu}(\omega_{0}))) t\}. 
\label{eqn:S24}
\end{eqnarray}

To obtain the first moment for the \(j\)-th interval of a step-like change of environmental parameter, 
we define the supermatrix during \(j\)-th interval over \(t_{j-1}  \leq t < t_{j} \) as
\begin{eqnarray}
\Xi^{\chi}_{d,M}(j) = - \sum_{\nu=L,R} \left[ {\begin{array}{*{20}{c}}
{k^{\nu}_{d}(j)}&{-k^{\nu}_{u}(j) e^{i \chi_{\nu} \hbar \omega_{0}}}\\
{-k^{\nu}_{d}(j) e^{-i \chi_{\nu} \hbar \omega_{0}}}&{k^{\nu}_{u}(j)}
\end{array}} \right], 
\nonumber 
\end{eqnarray}
where we set \(k^{\nu}_{d}(j)=\Gamma_{\nu} N_{\nu}(j)\) and \(k^{\nu}_{u}(j)=\Gamma_{\nu} (1+ N_{\nu}(j))\) with \(N_{\nu}(j)=1/(\exp[\hbar \beta_{j}^{\nu} \omega_{0}]-1)\) for the inverse temperature \(\beta_{j}^{\nu}\) of the \(\nu\)th environment during the \(j\)th interval. 
 
Using this definition, we obtain \(\langle \Delta q^{\nu} \rangle^{j}_{\delta t}\) as
\begin{equation}
\langle \Delta q^{\nu} \rangle^{j}_{\delta t} =\hbar \omega_{0} \{ A^{\nu}(j) \int_{0}^{\delta t} dt' \rho_{00}(t') - B^{\nu}(j) \delta t \}, 
\label{eqn:S25}
\end{equation}
with \(A^{\nu}(j) =-(k^{\nu}_{d}(j)+k^{\nu}_{u}(j))\) and \(B^{\nu}(j)= -k^{\nu}_{u}(j)\).

For the later convenience, let us rewrite the adiabatic contribution of \({\tilde J}^{R} \) with using Eqs.~(\ref{eqn:11}) and (\ref{eqn:12}) as
\begin{eqnarray}
\cg_{1}^{R}&=& \sum_{j=1}^{n} (A^{\nu}(j) \rho_{s}(j)-B^{\nu}(j)) \delta t \nonumber \\
&&\hspace{-0.5cm}  =\sum_{j=1}^{n} \frac{\Gamma_{L}\Gamma_{R} (N_{L}(j) - N_{R}(j))}{\sum_{\nu=L,R} \Gamma_{\nu} (1+2 N_{\nu}(j))} \delta t ,
\label{eqn:S26} \\
\cg_{2}^{R}&=& \sum_{j=1}^{n-1} \frac{A^{R}(j+1)}{\Lambda (j+1)}(\rho_{s}(j+1)-\rho_{s}(j) ) \nonumber \\
&&\hspace{-0.5cm} =\sum_{j=1}^{n-1} \frac{\Gamma_{R} (1+2 N_{R}(j+1))}{\sum_{\nu=L,R} \Gamma_{\nu} (1+2 N_{\nu}(j))}\nonumber \\
&&\hspace{-0.5cm} \times(\frac{\sum_{\nu=L,R}\Gamma_{\nu} (1+N_{\nu}(j+1))}{\sum_{\nu=L,R} \Gamma_{\nu} (1+2 N_{\nu}(j+1))}\nonumber \\
&&\hspace{0.5cm} -\frac{\sum_{\nu=L,R}\Gamma_{\nu} (1+N_{\nu}(j))}{\sum_{\nu=L,R} \Gamma_{\nu} (1+2 N_{\nu}(j))}).
\label{eqn:S27} 
\end{eqnarray}

In Appendix D, we show  the correspondences of \(\cg_{1}^{R}\) and \(\cg_{2}^{R}\) with the dynamical and geometric phases obtained under adiabatic approximation in \cite{Ren}.  We also find that \(\cg_{1}^{R}\)  for a single setting of environmental temperature corresponds to the current \(<J>\) obtained in \cite{Nicolin}.

\section{A TREATMENT FOR CONTINUOUS DRIVING PROTOCOL}
In this Appendix, we show that \(\cg_{1}^{\nu}\) and \(\cg_{2}^{\nu}\) obtained in this paper reduce to the dynamical and geometrical phases for continuous control in Ref.\cite{Ren}, respectively.  First, we assume that the relevant system can instantaneously follow the continuous driving of the environmental temperatures and the Markovian (long time) limit is reasonable at each time, which means that \(\Xi^{\chi}_{d,M}\) depends on time as
\begin{equation}
\Xi^{\chi}_{d,M}(t) =  \left[ {\begin{array}{*{20}{c}}
{\xi_{1}(t)}&{\xi^{\chi}_{2}(t)}\\
{\xi^{\chi}_{3}(t)}&{\xi_{4}(t)}
\end{array}} \right], 
\label{eqn:S28}
\end{equation}
where \(\xi_{m}(t)\) (\(m=1 \sim 4\)) are described by replacing \(\beta_{\nu}\) in Eq.~(\ref{eqn:S19}) with time dependent function corresponding to the driving protocol. In order to certify this situation, the correlation time of the environment is necessary to be much shorter than the relaxation time of the relevant system and the period of modulation.  We note that the time dependence in this Appendix  is different from the one for the non-Markovian case where the time dependence is determined by the system-environment interaction as in Eq.~(\ref{eqn:S10}).

We define the eigenvalues of \(\Xi^{\chi}_{d,M}(t) \) as \(\Lambda^{\chi}_{n,M}(t)\), and the right and left eigenvectors as \(|\lambda_{n,M}^{\chi}(t) \rangle\) and \(\langle \ell_{n,M}^{\chi}(t) |\) for \(n=0,1\), which satisfy the identity relation as \(\sum_{n=0,1} |\lambda_{n,M}^{\chi} (t) \rangle \langle \ell_{n,M}^{\chi}(t)|=1\). In the following, we abbreviate \(M\) in the eigenvalues and eigenvectors.  Taking the adiabatic approximation, we consider that the system instantaneously approaches to the stationary state, which corresponds to the eigenvector with zero eigenvalue. We assign the case to \(n=0\) where the eigenvalue and eigenvector are obtained as
\begin{eqnarray}
\Lambda^{\chi}_{0}(t) &=& \frac{\xi_{1}(t)+\xi_{4}(t)}{2}+\sqrt{(\frac{\xi_{1}(t)-\xi_{4}(t)}{2})^2 +\xi^{\chi}_{2}(t) \xi^{\chi}_{3}(t)},  
\nonumber \\ \label{eqn:S29}\\
|\lambda_{0}^{\chi}(t) \rangle &=& \frac{1}{F^{\chi}(t)} 
\left[ {\begin{array}{*{20}{c}}
{1}\\
{-\frac{G^{\chi}(t)}{\xi^{\chi}_{2}(t)}}
\end{array}} \right], \;
\langle \ell_{0}^{\chi}(t) | = [1,  -\frac{G^{\chi}(t)}{\xi^{\chi}_{3}(t)}],
\label{eqn:S30}
\end{eqnarray}
with
\begin{eqnarray}
F^{\chi}(t) &=& 1+ \frac{G^{\chi}(t)^2}{\xi^{\chi}_{2}(t) \xi^{\chi}_{3}(t)}, 
\label{eqn:S31}\\
G^{\chi}(t) &=& \frac{\xi_{1}(t)-\xi_{4}(t)}{2} - \sqrt{(\frac{\xi_{1}(t)-\xi_{4}(t)}{2})^2 +\xi^{\chi}_{2}(t) \xi^{\chi}_{3}(t)}. 
\nonumber \\\label{eqn:S32}
\end{eqnarray}
Using Eqs.~(\ref{eqn:S19}) and ~(\ref{eqn:S28}), we find the relations \(\xi^{\chi=0}_{2}(t)=-\xi_{4}(t)\) and \(\xi^{\chi=0}_{3}(t)=-\xi_{1}(t)\)  to give \(\Lambda^{\chi=0}_{0}(t)=0\). 

In order to discuss the adiabatic approximation in the quantum pumping, we use the procedure as in \cite{SN071,Ren} where we firstly divide the cycle of environmental parameter change into \(\delta t (\equiv \ct/n)\) and then take the Riemann sum.  Assuming that the system quickly approaches to the eigenstates \(|\lambda_{0}^{\chi}(t) \rangle\) and \(\langle \ell_{0}^{\chi}(t) |\), we insert the identity solution in each time step. In each division \(\delta t\), we set \(\kappa_{j}^{\chi}(\delta t)=\exp[\Xi^{\chi}_{d}(t_{j}) \delta t]  \) for the \(j\)th setting of  environmental parameters, which gives
\begin{eqnarray}
J =  \Bigl[\langle 1 | \frac{\partial}{\partial (i \chi)} \rho (\chi,\ct) \rangle \Bigl]_{\chi=0} \nonumber \\
&&\hspace{-4.5cm} \approx \Bigl[\langle 1 | \frac{\partial}{\partial (i \chi)} \prod_{j=1}^{n} \kappa_{j}^{\chi}(\delta t) | \rho (\chi,0) \rangle  \Bigl]_{\chi_{\nu}=0} \nonumber \\
&&\hspace{-4.5cm} \approx  \Bigl[ \langle 1 |  \frac{\partial}{\partial (i \chi)} \big\{ \exp[\sum_{j=1}^{n} \Lambda_{0}^{\chi}(t) \delta t] \nonumber \\
&&\hspace{-4cm} \times \exp[-\sum_{j=1}^{n} \langle \ell_{0}^{\chi}(t) | {\dot \lambda}_{0}^{\chi}(t) \rangle \delta t] | \lambda_{0}^{\chi}(t_{n}) \rangle \big\}  \Bigl]_{\chi_{\nu}=0}, 
\label{eqn:S33}
\end{eqnarray}
where we define \(| {\dot \lambda}_{0}^{\chi}(t) \rangle = | \frac{\partial}{\partial t} \lambda_{0}^{\chi}(t) \rangle\), and use the relations as \(\langle \ell_{0}^{\chi=0}(t)|= \langle 1|\) and \(\langle 1| \lambda_{0}^{\chi=0}(t) \rangle =1\).
Taking the Riemann sum to describe the quantum pumping for continuous driving protocol, we obtain the transferred quantity between the \(\nu\)th environment and the relevant system as 
\begin{eqnarray}
{\tilde J}^{\nu}\nonumber \\
&&\hspace{-1cm} =\frac{1}{\ct}  \int_{0}^{\ct} dt' \Bigl[\{\frac{\partial}{\partial (i \chi_{\nu})} \{\lambda^{\chi}_{0}(t')\}   -\langle \frac{\partial}{\partial (i \chi)} \ell_{0}^{\chi}(t') |{\dot \lambda}_{0}^{\chi}(t') \rangle \} \Bigl]_{\chi_{\nu}=0} \nonumber \\
&&\hspace{-1cm} \equiv  J ^{\nu}_{1}+J^{\nu}_{2},
\label{eqn:S34}
\end{eqnarray}
where we set the counting field between the \(\nu\)th environment and the relevant system as \(\chi_{\nu}\).
Substitution of Eqs.~(\ref{eqn:S29}) --~(\ref{eqn:S32}) into \(J ^{\nu}_{1}\) gives
\begin{eqnarray}
J ^{\nu}_{1} &=& -\frac{1}{\ct} \int_{0}^{\ct} dt' \frac{\xi_{1}(t') \eta^{\nu}_{2}(t') + \eta^{\nu}_{3}(t') \xi_{4}(t')}{\xi_{1}(t')+\xi_{4}(t')}, 
\label{eqn:S35}
\end{eqnarray}
with \(\eta^{\nu}_{m}(t)=-\Big[\frac{\partial}{\partial (i \chi_{\nu})} \xi^{\chi_{\nu}}_{m}(t) \Big]_{\chi_{\nu}=0}\) for \(m=2,3\). When we use the elements in Eq.~(\ref{eqn:S19}), we obtain the transferred quantity between the \(R\)th environment and the relevant system as 
\begin{equation}
J^{R}_{1}=  \frac{\hbar \omega_{0}}{\ct} \int_{0}^{\ct} dt' \frac{\Gamma_{L} \Gamma_{R}(N_{L}(t')-N_{R}(t'))}{K}, 
\label{eqn:S36}
\end{equation}
with \(K \equiv -(\xi_{1}(t')+\xi_{4}(t'))=\sum_{\nu=L,R} \Gamma_{\nu} (1+2 N_{\nu}(t'))\). Taking the limit of \(n \rightarrow \infty\) and \(\delta t \rightarrow 0\) in Eq.~(\ref{eqn:S26}), we find that \(\frac{\hbar \omega_{0}}{\ct} \cg^{R}_{1}\) coincides with \(J^{R}_{1}\) given by Eq.~(\ref{eqn:S36}) and that it also coincides with the dynamical phase part called as \(J_{{\rm dyn}}\) in Ref.\cite{Ren}.  This means that Eq.~(\ref{eqn:S26}) corresponds to the expression of the dynamical phase part in Ref.\cite{Ren} under the piecewise temperature control.

Similarly, we can rewrite \(J^{\nu}_{2}\) in Eq.~(\ref{eqn:S34}) in the form as 
\begin{equation}
J^{\nu}_{2} = \frac{1}{\ct} \int_{0}^{\ct} dt' \frac{(\eta^{\nu}_{2}(t') - \eta^{\nu}_{3}(t'))}{\xi_{1}(t')+\xi_{4}(t')} \frac{d}{dt'} \frac{\xi_{4}(t')}{\xi_{1}(t')+\xi_{4}(t')},
\label{eqn:S37}
\end{equation}
which corresponds to the Riemann sum of Eq.~(\ref{eqn:S27}) by taking \(n \rightarrow \infty\) and \(\delta t \rightarrow 0\).
Using the fact that \(T_{\nu}\) is time dependent under the continuous driving, Eq.~(\ref{eqn:S37})  is rewritten as
\begin{eqnarray}
J^{\nu}_{2} &=& - \frac{1}{\ct} \int_{0}^{\ct} dt' \frac{(\eta^{\nu}_{2}(t') - \eta^{\nu}_{3}(t'))  (\xi_{1}(t')-\xi_{4}(t'))}{(\xi_{1}(t')+\xi_{4}(t'))^3} \nonumber \\
&&\hspace{1.5cm} \times ( \frac{d\xi_{1}}{dT_{L}} \frac{dT_{L}}{dt'} +\frac{d\xi_{4}}{dT_{R}} \frac{dT_{R}}{dt'}), \nonumber \\ 
\label{eqn:S38}
\end{eqnarray}
where we use the relation \(\frac{d\xi_{1}}{dT_{\nu}} =\frac{d\xi_{4}}{dT_{\nu}} \).  Application of the Green's theorem which transforms the line integral in Eq.~(\ref{eqn:S38}) into the surface integral over the temperature variable \(T_{L}\) and \(T_{R}\) gives
\begin{eqnarray}
J^{\nu}_{2} \nonumber \\
&&\hspace{-1cm}= - \int \int d{T_{L}} d{T_{R}} \Bigl\{\frac{d}{dT_{R}}\Bigl(\frac{(\eta^{\nu}_{2} - \eta^{\nu}_{3}) (\xi_{1}-\xi_{4})}{(\xi_{1}+\xi_{4})^3} \frac{d\xi_{1}}{dT_{L}} \Bigl)\nonumber \\
&&\hspace{1.5cm}-\frac{d}{dT_{L}}\Bigl(\frac{(\eta^{\nu}_{2} - \eta^{\nu}_{3}) (\xi_{1}-\xi_{4})}{(\xi_{1}+\xi_{4})^3} \frac{d\xi_{4}}{dT_{R}} \Bigl) \Bigl\}. \nonumber \\
\label{eqn:S39}
\end{eqnarray}
For the right counting field, we obtain
\begin{equation}
J^{R}_{2} =\frac{\hbar \omega_{0}}{\ct}  \int \int d{T_{L}} d{T_{R}} \Bigl\{\frac{2 \Gamma_{L} \Gamma_{R} (\Gamma_{L}+\Gamma_{R})}{K^3} \frac{d n_{R}}{dT_{R}} \frac{d n_{L}}{dT_{L}}\Bigl\} , \\
\label{eqn:S40}
\end{equation}
which coincides with the geometrical phase part of the current \(J_{{\rm geom}}\) in Ref.\cite{Ren} by denoting \(C_{\nu} \equiv -\frac{d n_{\nu}}{dT_{\nu}}\) for \(\nu=L,R\).  From  Eqs.~(\ref{eqn:S37}) and ~(\ref{eqn:S40}), we find that  Eq.~(\ref{eqn:S27}) corresponds to the expression of the geometrical phase part in Ref.\cite{Ren} under the piecewise temperature control.

\section{ON THE PARAMETER SETTINGS FOR NUMERICAL EVALUATION}
Let us discuss about the parameter setting of \(\omega_{c}=3 \omega_{0}\) used in the numerical evaluations.  Since we use the expressions for the Markovian (long time) approximation, it is necessary to discuss how the value of  \(\omega_{c}\) is consistent with the approximation.  The most important feature of the Markovian approximation appears in the decay constant \(\Lambda (j)\) in Eq.~(9).  We focus on a single setting of environmental temperature in the following and abbreviate \(j\) in this Appendix.  The definition of \(\Lambda \) is given by \(\Lambda =\sum_{\nu}A^{\nu}\) with \(A^{\nu}=-\Gamma_{\nu} (1+2 n_{\nu}(\omega_{0}))\). The constant \(\Lambda \) represents one of the eigenvalue of \(\Xi^{\chi=0}_{d,M}\).  Defining the elements of \(\Xi^{\chi=0}_{d,M}\) as 
\begin{eqnarray}
\Xi^{\chi=0}_{d,M} &=& - \sum_{\nu=L,R} \left[ {\begin{array}{*{20}{c}}
{\Gamma_{\nu} n_{\nu}(\omega_{0})}&{-\Gamma_{\nu} (1+n_{\nu}(\omega_{0})) }\\
{-\Gamma_{\nu} n_{\nu}(\omega_{0}) }&{\Gamma_{\nu} (1+n_{\nu}(\omega_{0}))}
\end{array}} \right]  \nonumber \\
&&\hspace{-0.5cm}=\left[ {\begin{array}{*{20}{c}}
{a_{1}}&{-a_{2}}\\
{-a_{1}}&{a_{2}}
\end{array}} \right] , 
\label{eqn:S41}
\end{eqnarray}
we find that the expression of \(\Lambda \) is rewritten as \(\Lambda =a_{1}+a_{2}\).  We define the value as \(\Lambda_{S}\) in this Appendix.  Before taking the Markovian approximation, the corresponding supermatrix is given by
\begin{equation}
\Xi^{\chi=0}_{d}(t)= 
\left[ {\begin{array}{*{20}{c}}
{a_{1}(t)}&{-a_{2}(t)}\\
{-a_{1}(t)}&{a_{2}(t)}
\end{array}} \right] ,
\label{eqn:S42} 
\end{equation}
with \(a_{1}(t)=- \int_{0}^{t} d\tau V_{+}(\tau)\) and  \(a_{2}(t)=- \int_{0}^{t} d\tau V_{-}(\tau)\).  In obtaining  Eq.~(\ref{eqn:S42}), we use the relation as \(W_{\pm}^{\chi=0}(\tau)=-V_{\mp}(\tau)\) which is obtained by setting \(\chi=0\) in Eq.~(\ref{eqn:S12}) and comparing it with Eq.~(\ref{eqn:S11}).  From Eq.~(\ref{eqn:S42}), we obtain the expression of \(\Lambda \) before taking the Markovian approximation as \(\Lambda (t)=a_{1}(t)+a_{2}(t)\).  

We show the time dependence of \(\Lambda (t)\) for scaled time variable as \({\tilde t}=\omega_{0} t\) with changing  \(T_{L}\) and \(T_{R}\) in Fig.3.  We set \(s=0.01\), \(\omega_{c} =3 \omega_{0}\), and \(\hbar \omega_{0}=25 {\rm meV}\)  which are the same as used in Fig.1.  The temperature settings are (1) \(T_{L}=100 (K), T_{R}=200(K)\) , (2) \(T_{L}=200 (K), T_{R}=300(K)\), and (3) \(T_{L}=T_{L}(0), T_{R}=T_{R}(0)\) which represents typical three points on the circle of modulation \(T_{L}(t)= 200 + 100 \cos(\omega_{0} t+\pi/4)\), and \(T_{R}(t)= 200 + 100 \sin(\omega_{0} t+\pi/4)\).  Dashed lines show the values \(\Lambda_{S}\) under the Markovian approximation for each setting.  We find that \(\Lambda (t)\) asymptotically approaches to \(\Lambda_{S}\). When we set \(\hbar \omega_{0}=25 {\rm meV}\), the unit of the scaled time axis corresponds to  \(0.026 (ps)\). 
\begin{figure}[h]
\begin{center}
\hspace{0cm}
\includegraphics[scale=0.6]{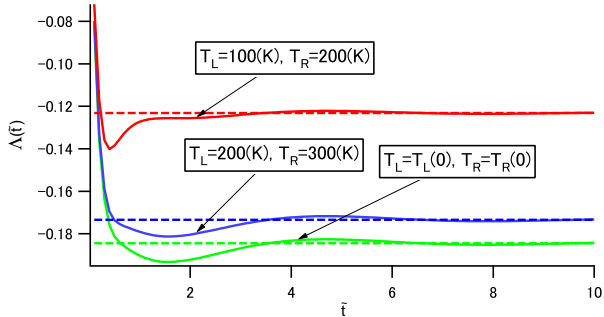}
\end{center}
\caption{(Color online) Time dependence of \(\Lambda({\tilde t})\) for \(s=0.01\), \(\omega_{c} =3 \omega_{0}\), and \(\hbar \omega_{0}=25 {\rm meV}\) with changing \(T_{L}\) and \(T_{R}\): (1) the red (lighter gray) line corresponds to \(T_{L}=100(K), T_{R}=200(K)\), (2) the blue (dark gray) line to \(T_{L}=200(K), T_{R}=300(K)\), and (3) the green (lightest gray) line to \(T_{L}=T_{L}(0), T_{R}=T_{R}(0)\). The dashed lines represent the values under the Markovian (long-time) approximation for the respective cases.
}
\label{fig:fig3}
\end{figure}

In fig.4, we show time dependence of relative error, \((\Lambda ({\tilde t})-\Lambda_{S})/\Lambda_{S}\).  Considering that the time interval \(\delta t \) is \(\delta t \approx  0.031(ps)\) for \(n=41\) divisions in the case of \(\Omega=5 {\rm THz}\), the Markovian approximation is found to be within 5 \(\%\) error at least at the end of each \(\delta t\) for \(5 {\rm THz}\).  From this figure, we find that the non-Markovian effect has larger contribution as the frequency of temperature modulation becomes larger.  This means the necessity to include the non-Markovian  effect, which will be treated in our future study.   
\begin{figure}[h]
\begin{center}
\hspace{0cm}
\includegraphics[scale=0.6]{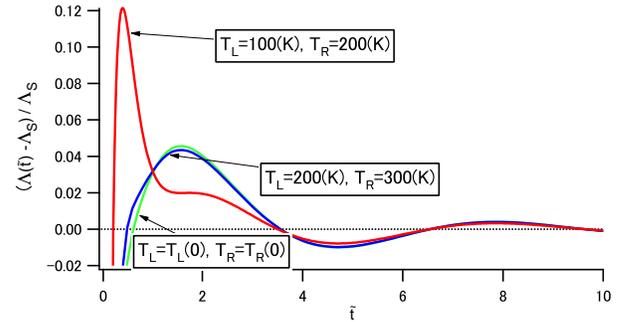}
\end{center}
\caption{(Color online) Time dependence of the relative error, \((\Lambda ({\tilde t})-\Lambda_{S})/\Lambda_{S}\). Parameters and color markings are the same as in Fig. 3. }
\label{fig:fig4}
\end{figure}


\begin{thebibliography}{99}
\bibitem{ucom} The master equation obtained in \cite{EHM} can be rederived by using the projection operator method\cite{US} and modifying the Liouville operator describing the time evolution of the total system to include the counting field: \(\elc A=\frac{1}{\hbar} [\ch^{\chi}A-A\ch^{-\chi}] \equiv \frac{1}{\hbar} [\ch,A]_{\chi}\) for an arbitrary operator \(A\) with \(\ch^{\chi}=e^{(i/2) \chi \ch_{E}} \ch e^{-(i/2) \chi \ch_{E}}=\ch_{0}+\ch_{1}^{\chi}\).
\end{thebibliography}
\end{document}